\documentclass[conference]{IEEEtran}
\usepackage[utf8]{inputenc}
\usepackage[english]{babel}
\usepackage{amsmath}
\usepackage{amsfonts}
\usepackage{amssymb}
\usepackage{graphicx}

\usepackage{color}
\usepackage{xspace}
\usepackage[normalem]{ulem}
\usepackage{hyperref} 
\usepackage{setspace}

\makeatletter\def\@IEEEpubidpullup{8\baselineskip}\makeatother

\begin{document}

\IEEEoverridecommandlockouts
\IEEEpubid{
\parbox{\columnwidth}{\vspace{-4\baselineskip}
\textbf{IEEE Copyright Notice\\
\copyright\space2021 IEEE. Personal use of this material is permitted. Permission from IEEE must be obtained for all other uses, in any current or future media, including reprinting/republishing this material for advertising or promotional purposes, creating new collective works, for resale or redistribution to servers or lists, or reuse of any copyrighted component of this work in other works.}
\hfill\vspace{-0.8\baselineskip}\\
\begin{spacing}{1.2}
\small\textit{Accepted to IEEE I$^2$MTC – International Instrumentation and Measurement Technology Conference, May 17-20, 2021}
\end{spacing}
\hfill}\hspace{0.9\columnsep}\makebox[\columnwidth]{\hfill}}
\IEEEpubidadjcol

\title{\textbf{Frequency span optimization for asymmetric resonance curve fitting}}

\author{Kostiantyn Torokhtii, Andrea Alimenti, Nicola Pompeo, Enrico Silva \\
\textit{Dipartimento di Ingegneria, Universit\`a Roma Tre, Roma, Italy} \\
kostiantyn.torokhtii@uniroma3.it}
\date{}

\maketitle

\begin{abstract}
The wide application of the modern resonant measurement techniques makes all the steps of the measuring process, including data acquisition more efficient and reliable. 
Here we investigate the multidimensional space of the parameters to determine the optimum span for resonant measurements. 
The study concentrated on experimental systems with standard performance and capabilities. 
We determine the range of the optimum span for the resonant frequency and quality factor by simulating and fitting resonant curves with different levels of asymmetry.

\end{abstract}

\section{Introduction}
Measurements at microwaves have been for a long time a prerogative of specialized laboratories. 
Lastly, this type of measurement has become more accessible also for non-specialists. 
%\textoutr{Thanks to the interest to high frequencies,} 
Many devices operating at GHz have become available for a wide range of applications \cite{Swicord08}. 
Enormous interest in the new communication technologies operating at GHz increased the study of the characteristics of devices and materials at microwaves.

There are many techniques for material characterization at microwaves \cite{ChenBook}.
The resonant method is one of the widely used in applications where precision and sensitivity are required from fundamental physics to industrial applications. 
Well-known applications are dielectric permittivity characterization \cite{Chen18, Schultz18}, temperature sensing \cite{Adiyan19} and (super)conductor surface impedance measurements \cite{IEC, AlimentiTAS19, AlimentiMST19, PompeoMeas17}. 
This widespread use of the resonant methods pushes to optimize both the measurement setup and the measurement procedure.

The highest precision is reached implementing the resonant measurement technique with a Vector Network Analyzer (VNA). 
Progress in the hardware operating at GHz frequencies opened the possibility to create more portable stand-alone measurement devices \cite{You17}. 
In the last years a wide range of low-cost VNAs became available on the market \cite{Verhaevert17, Will08}. 
Another low price solution is represented by scalar network analyzers.

Low-cost VNAs have limited performance and functionality, which determine often a low number of measured frequency points and/or long acquisition time. 
In these cases, one of the crucial setup choices of the resonant system becomes the optimal acquisition of the resonant curve measured against the frequency. 

For ideal resonant systems, the resonance response in frequency has a Lorentzian curve shape \cite{Petersan98}, whereas 
for a real resonator the resonance curve can become asymmetric and non-Lorentzian. 
Once a real resonator is inserted in the measurement chain, inaccurate data could affect the final results. 
In order to attenuate the issue, some a priori knowledge regarding the source of asymmetry needs to be taken into account. 

{We addressed previously the relevance of the calibration on the uncertainty of the measured parameters (quality factor and resonant frequency) of resonators \cite{Torokhtii18}. }
In this article, we focus instead on the optimization of the frequency range of a resonance curve.
We investigate the causes leading to the asymmetry of the resonance curve, exploring the corresponding parameter space through a series of simulations in order to optimize the measurement process.
We note that the following discussion is valid also for other applications different from microwaves, such as spectroscopy, where parameters must be extracted from Lorentzian-like curves.

\section{Resonant measurements}
Measurements of the resonance curve at microwaves can be performed resorting to both vector and scalar network analyzers. 
In the second case, no information regarding the phase can be extracted.
In both cases, the measured quantities are the well-known scattering parameters ($S$-parameters) \cite{ChenBook}. 
 
In the following, we focus on transmission measurements, so that the relevant scattering coefficient is $S_{21}$ or $S_{12}$ which we define here as $S_{tr}$.
Once measured vs the frequency $f$, the scattering coefficient $S_{tr}$ yields a resonant curve which, in the ideal case, has the following expression \cite{Pompeo17}:

\begin{equation}
S_{tr}(f)=\frac{S_{tr}(f_0)}{1+2i Q\frac{f-f_{0}}{f_{0}}}
\label{eq:S21ideal}
%\vspace{-1 ex}
\end{equation}

whose square modulus has a Lorentzian shape.

In this ideal case, the determination of the resonant frequency $f_0$ and of the resonator quality factor $Q$ is straightforward with standard approaches. 
$f_0$ and $Q$ are the parameters of interest for such applications as a resonant thermometer, where $f_0$ is used for temperature sensing \cite{Adiyan19}, and surface impedance $Z_s$ measurements, where based on $Q$ and $f_0$ the real and imaginary parts of $Z_s$ can be extracted correspondingly  \cite{PompeoMSR14, PompeoTAS18}. 

In real-world measurement systems, one often faces the situation where a non-ideal curve must be properly interpreted to extract the desired quantities. 
When knowledge about the causes of the non-idealities are available, it is possible to
create a model of the whole measuring system with well-defined characteristics of all the components, so that a calibration can be performed to correct the resonance response.

However, the {end} user of the measuring device {may} not have the technical expertise or possibilities to intervene in such particular issues. 
It is then of interest to address the problem 
from a general point of view without reference to a specific setup, i.e. by providing a general approach independent from the specific causes of the non-idealities. 
In doing so, we put a particular attention on the estimation of the uncertainties on the resonant system parameters $Q$ and $f_0$. 

The simplest method for the determination of $Q$ and $f_0$ is the so-called  ``-3dB method'',
where $f_0$ is determined as the frequency of the maximum of the transmission coefficient, and the -3dB level with respect to this maximum allows to compute the so-called full-width half maximum (FWHM) frequency range $\Delta f_{FWHM}$ hence $Q=f_0/\Delta f_{FWHM}$ \cite{Petersan98}. 

This type of estimation is widely used in a large part of software applications including internal VNAs programs. 
In these cases, the uncertainty estimation is highly dependent on the number of points in frequency and on the noise level.

A more accurate way to extract $f_0$ and $Q$ is to fit the resonant curve vs frequency.
When non-idealities are present, the use of the ideal Lorentz model yields results which can be quite inaccurate.
Previous studies in which different resonant curve fitting models {were compared} have shown a good effectiveness of the Fano resonance model \cite{Pompeo17, Yoon13}.
{In particular, the feasibility of uncalibrated measurements was explored \cite{Torokhtii18}.}

{Here, resorting to the already-demonstrated robust Fano resonance model, we proceed further in the optimization process by addressing the problem of the determination of the acquisition frequency span capable of yielding the minimum} uncertainty of the resonant quantities $Q$ and $f_0$.

In the following we first describe the numerical simulations and the results, and then we present experimental measurements aimed at the validation of the simulation study.

\section{Fitting procedure}
Focusing on the microwave measuring devices with limited performance, we consider the case of cheap scalar measurement systems, i.e. capable of measuring only the amplitude of the scattering coefficients, with a limited number $N$ of measured points. 

The Fano model for the transmission coefficient is based on the Lorentzian model with the addition of a constant complex parameter modelling a cross-coupling contribution.

The absolute value of the Fano expression is thus:

\begin{equation}
|S_{tr}(f)|=\left|\frac{S_{M}}{1+2i Q\frac{f-f_{0}}{f_{0}}}+{S}_c\right|
\label{eq:S21}
\vspace{-1 ex}
\end{equation}
where $f_0$ and $Q$ are resonant frequency and quality factor, $S_{M}$ = $S_{tr}(f_0)$, $S_c$ is the complex cross-coupling constant. 
The $S_c$ parameter controls the amount of ``non-ideality'' of the resulting curve, being capable to produce pure Lorentzian curves for $S_c=0$ to extremely deformed, asymmetric curves.

For the fit, we choose a standard implementation Levenberg–Marquardt algorithm \cite{Marquardt63} from \texttt{curvefit} module of python scipy package \cite{JonePy, More90MINPACK}. 
Through this implementation, the non-linear least squares problem could be solved. 
The numerical calculation of the Jacobian provides the covariance matrix from which the uncertainty of the fitting parameters can be estimated as square roots of variances.
Since the algorithm requires starting guess values for $Q$ and $f_0$, we determine them with the -3dB method.

Aiming at the optimization of the frequency interval to minimize the contribution of the fit on the uncertainty of the $f_0$ and $Q$, a series of asymmetrical curves was numerically simulated according to Eq. \eqref{eq:S21} with varying resonant parameters. Subsequently, {Gaussian noise with standard deviation $3\cdot10^{-4}$} was added and the curves have been fitted.

{The parameter space to be explored is chosen in order to enclose the parameters of a real dielectric resonators \cite{Torokhtii18, AlimentiMST19}, which has been subsequently used to validate the simulations (see Section \ref{sec:exp}), but sufficiently wide to provide generalized results.}

Since we are interested only on the contributions responsible for non-ideal curves parameters, other parameters could be fixed such as $S_{M}$, $f_0$ and noise level. 
We note that, in particular, the choice of $f_0$ is irrelevant since the frequency range width is normalized to the ratio $f_0/Q$ and no contributions from the frequency-dependent background were added. 
One of the main parameters responsible for the signal to noise ratio and consequently the fit quality is $Q$-factor. 
We select $Q$-factors between $2\cdot10^3$ — $10^5$ to cover the range of $Q$-factors of medium $Q$ cavities to high-$Q$ resonators \cite{ChenBook}.

\begin{figure}[htbp]
\centerline{\includegraphics[width=0.5\textwidth]{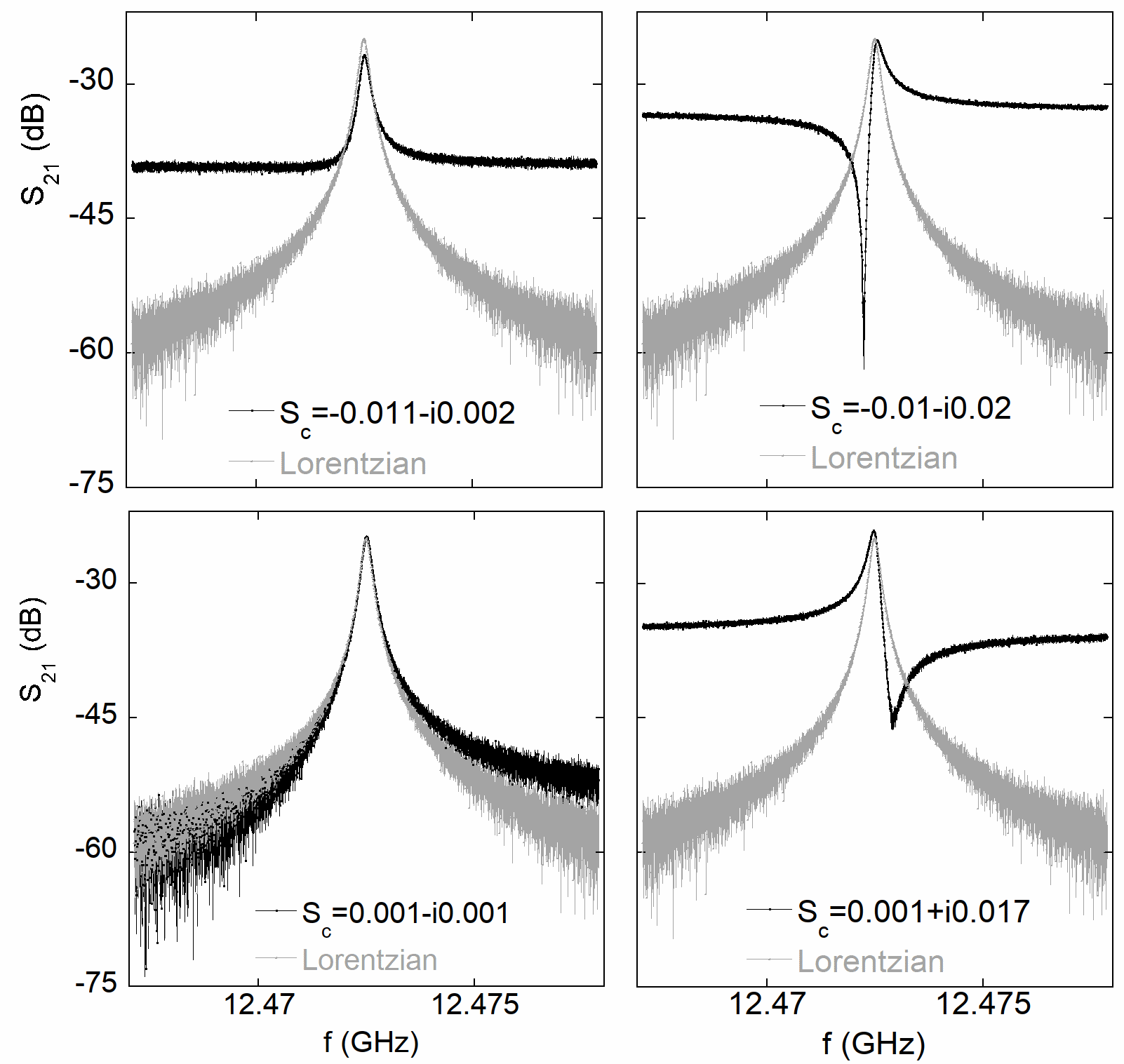}}
\caption{Asymmetric resonant curves simulated for different asymmetries at different $S_c$ and compared with simple Lorentzian. Normally distributed noise with a standard deviation $3\cdot10^{-4}$ has been added.}
\label{fig:rescurves}
\end{figure}

We keep a fixed number of data points equal to 1601 {as a good compromise between acquisition speed and performance. It should be noted that low-cost VNA even capable to acquire a large number of data-points (with maximum point number between 16001 to 1000001) are relatively slow in acquisition (with speed between 1ms/point to 130~$\mu$s/point) \cite{MS46122B, MegiQ, PicoVNA}.}
Concerning $S_c$, we determined that even with parameters $\vert Re(S_c)\vert \leqslant 0.012$ and $\vert Im(S_c)\vert \leqslant 0.03$, reproducing typical ``non-idealities'' up to extremely asymmetric curves, the ``-3db'' method could always be used to determine guess values of $f_0$ and $Q$.
In Fig. \ref{fig:rescurves} an example of the curves with low and high asymmetry is reported.

The determination of the optimum frequency span was performed in two steps. 
First, for each value of $S_c$ a series of curves was generated with different frequency span around $f_{0}$ in the frequency range $f_{0,set} \pm C_{FWHM} \Delta f_{FWHM,set}$, where $f_{0,set}$, $\Delta f_{FWHM,set}$ are the set nominal values and $C_{FWHM}$ is an integer multiplier constant.

It is clear that selecting a too wide span at a fixed number of the acquisition points can greatly raise the uncertainty of the determined parameters and lower the fit quality in general. Thus we limit $C_{FWHM}$ from 1 up to 20 obtaining maximum span equal to 20 FWHMs.

\begin{figure}[htbp]
\centerline{\includegraphics[width=0.5\textwidth]{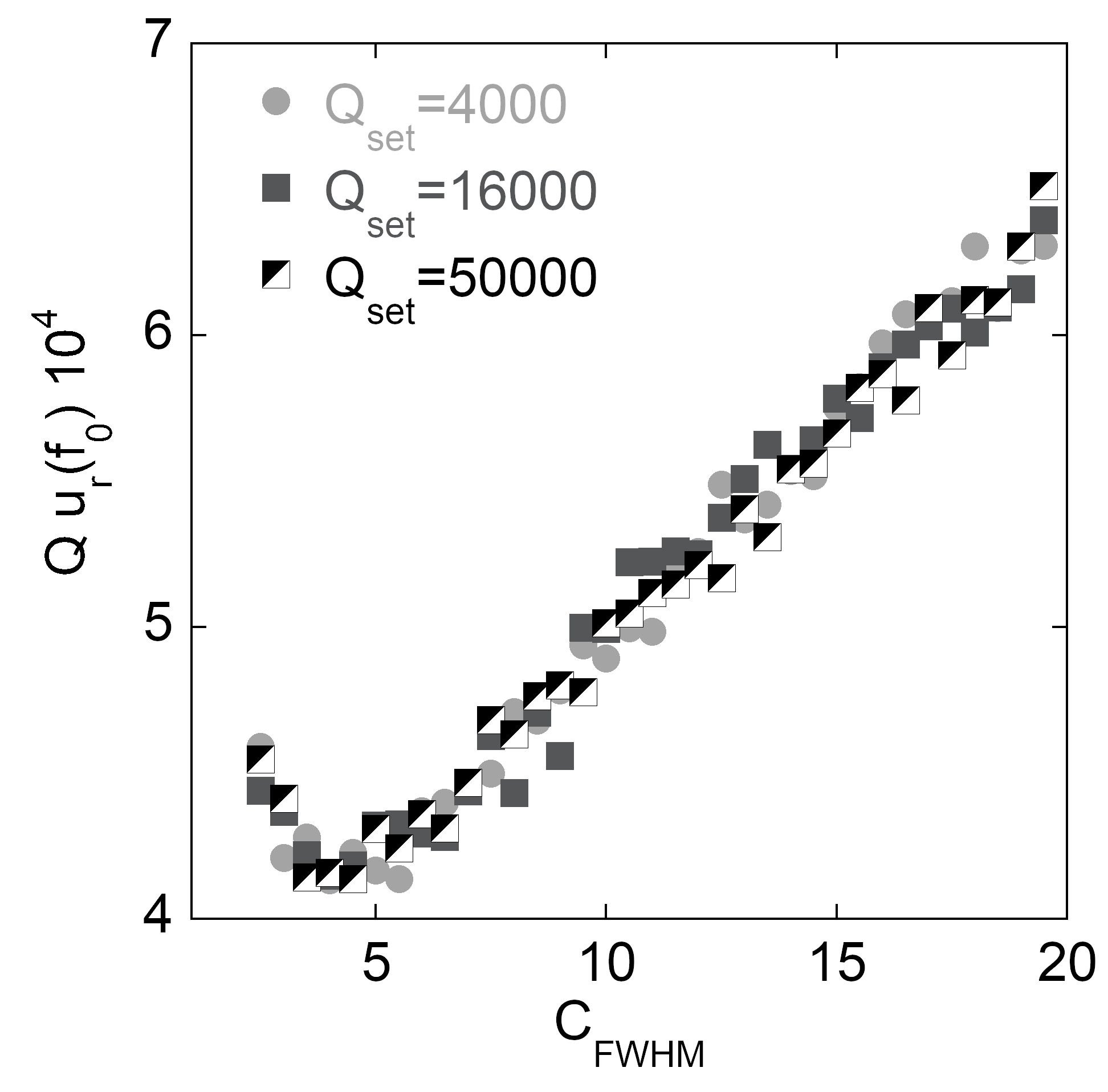}}
\caption{Example of the relative uncertainty on $f_{0}$ as a function of the frequency span for different $Q$-factors.}
\label{fig:urf0}
\end{figure}

\begin{figure}[htbp]
\centerline{\includegraphics[width=0.5\textwidth]{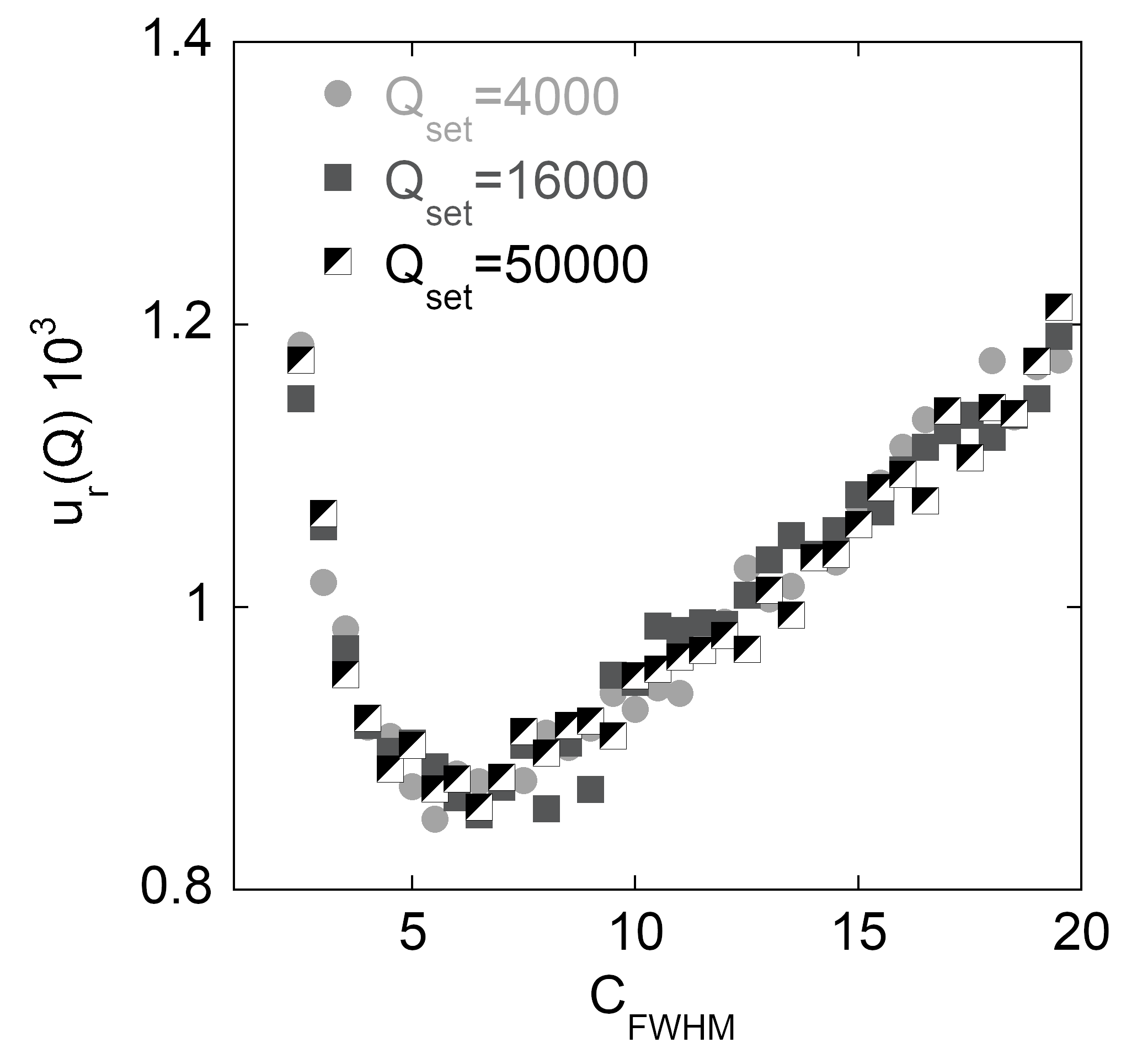}}
\caption{Example of the relative uncertainty on $Q$ as a function of the frequency span for different $Q$-factors.}
\label{fig:urQ}
\end{figure}

In Figs. \ref{fig:urf0} and \ref{fig:urQ} an example of the dependence of the relative uncertainties ($u_r(A)=u(a)/a$, where $a$ is the input estimate) to the variation of chosen span for \mbox{$S_c=(6-i 7)\cdot10^{-3}$} is reported.
We observe minima in both $u_r(f_{0})(C_{FWHM})$ and $u_r(Q)(C_{FWHM})$ dependencies, which we anticipate we observed also for other values of $S_c$.
In the second step, we determined the minimum of each curve by a polynomial fit.


\section{Simulation results}
An investigation of the whole space of the parameters consisting of variations of $S_c$ and $Q$ was performed by fitting more than 50000 curves. 

It is worth mentioning that a systematic offset between the set $f_{set}$ and fit $f_0$ resonant frequencies $ \dfrac{f_{set}-f_{0}}{f_{set}} $ decreases with growing $Q$ from $0.3\cdot10^{-6}$ to $6\cdot10^{-9}$. 
Moreover, in this large multidimensional space of parameters, a subset of the generated resonance curves were so distorted that no fit could be performed.
Thus, during the parametric sweeps, we isolated these cases discarding them from the statistics.

\begin{figure}[htbp]
\centerline{\includegraphics[width=0.5\textwidth]{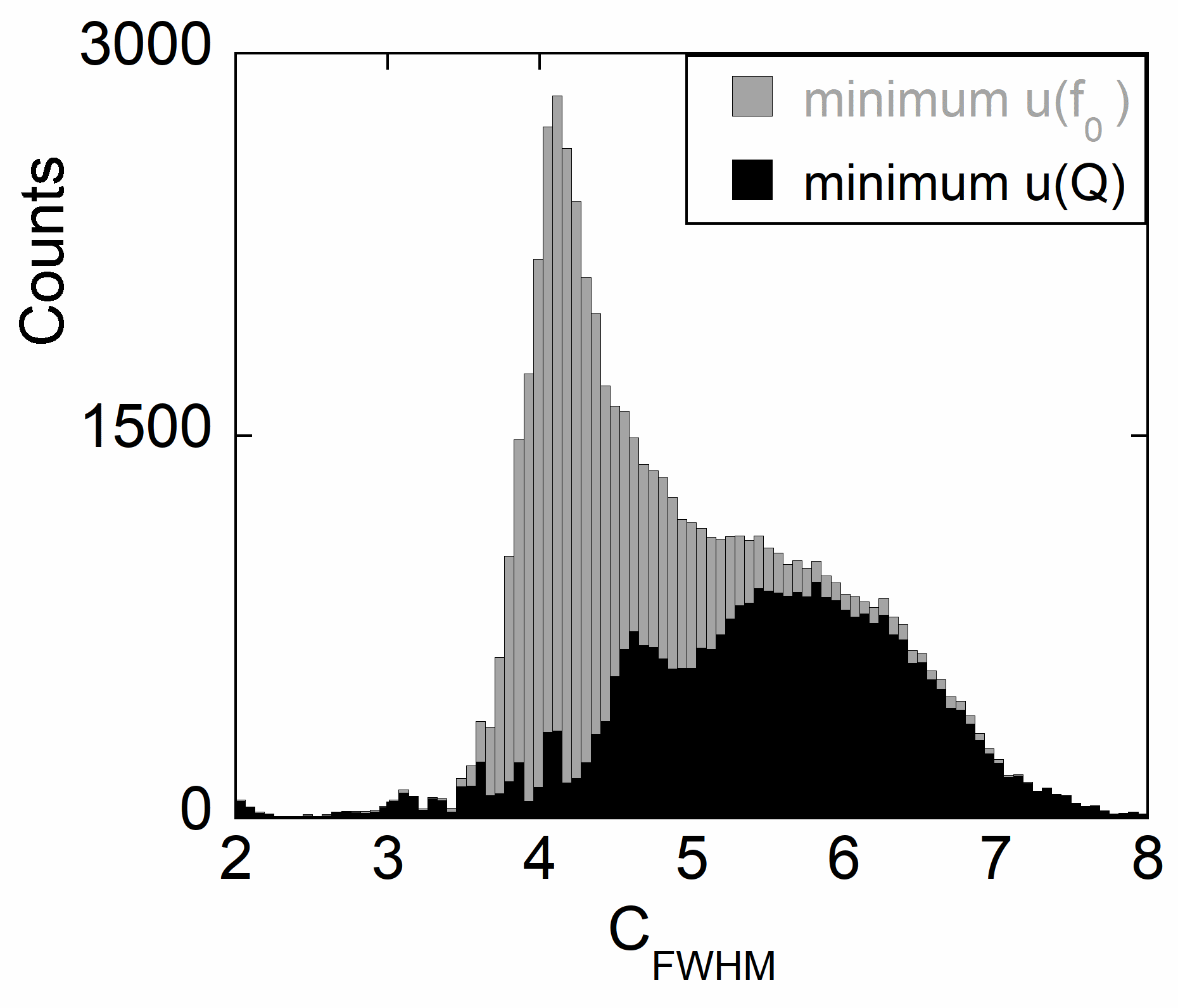}}
\caption{Histogram of position of the minimum of the relative uncertainty on $f_0$ and $Q$.}
\label{fig:muncdist}
\end{figure}

In Fig. \ref{fig:muncdist} we report the distribution, among all the valid runs with different asymmetries, of the frequency span width, in terms of the number $C_{FWHM}$ of FWHM spans, which yielded the
minimum relative uncertainty of $f_0$ and $Q$. 
On average, we obtain that the minimum for $u_r(f_0)$ and $u_r(Q)$ can be obtained with span width $(4.1 \pm 0.4) FWHM$ and $(6.1 \pm 0.9) FWHM$, respectively.

We also performed an additional study to determine the effect of the noise on the observed result. 
Increasing the noise by a factor of 2 does not alter the value of $C_{FWHM}$ corresponding to the minimum uncertainty. 
It should be noted that a larger noise level increases the rate of the failed fits to be discarded.

From our results, it is evident that for different applications of the resonant techniques a different optimal span should be selected to minimize the  uncertainty of the measured parameters. 
As an example, for measurements where the resonant frequency is the fundamental parameter, as in the measurements of the relative permittivity, the optimum span is 4.1 FWHM. 
By contrast, where the $Q$-factor is the desired measurement parameter (as in dielectric loss tangent measurements), the optimum span is higher, equal to 6.1 FWHM. 
More complex problems, such as the measurement of $Z_s$, require the simultaneous determination of both $f_0$ and $Q$, so that a tuning of the optimum frequency span within the 4.1 and 6.1 FWHMs range should be done, on the basis of the sensitivity and of the measurement conditions.

\section{Experiment}
\label{sec:exp}

An experimental study was performed to verify the results of the numerical simulations. 
For the measurements, a sapphire dielectric resonator (DR), designed and optimized for high repeatability and sensitivity, was used.
The resonator operates in transmission, using the  $TE_{011}$ mode with a resonant frequency $f_0=$12.9~GHz and an average room temperature $Q$-factor equal to 4500. 
The geometry of the resonator allows the selected $TE_{011}$ mode to be well separated in frequency from spurious modes, so excluding inter-mode interferences. 
More details about the structure and characteristics of such types of resonant cells can be found elsewhere \cite{PompeoMSR14}.

Measurements were performed by means of an Anritsu 37269D VNA. 
The DR is designed to allow the variation of the cross-coupling amount by varying the position of the coupling loops.
Thus, we were able to control manually the asymmetry of the resonant curve $S_{tr}$. 
By the variation of the coupling loops position, several resonant curves with different asymmetry were measured. 
For each cross-coupling setting, a series of measurements was performed as follows. 
Maintaining a fixed center frequency near the resonant frequency of $TE_{011}$, 
we performed acquisitions of the resonant curve for each selected frequency span, for $C_{FWHM}$ values between 1 and 20 in steps of 1.
Each curve was measured with a number of frequency points equal to 1601, the maximum available for the {Anritsu 37269D} VNA used here. 

\begin{figure}[htbp]
\centerline{\includegraphics[width=0.5\textwidth]{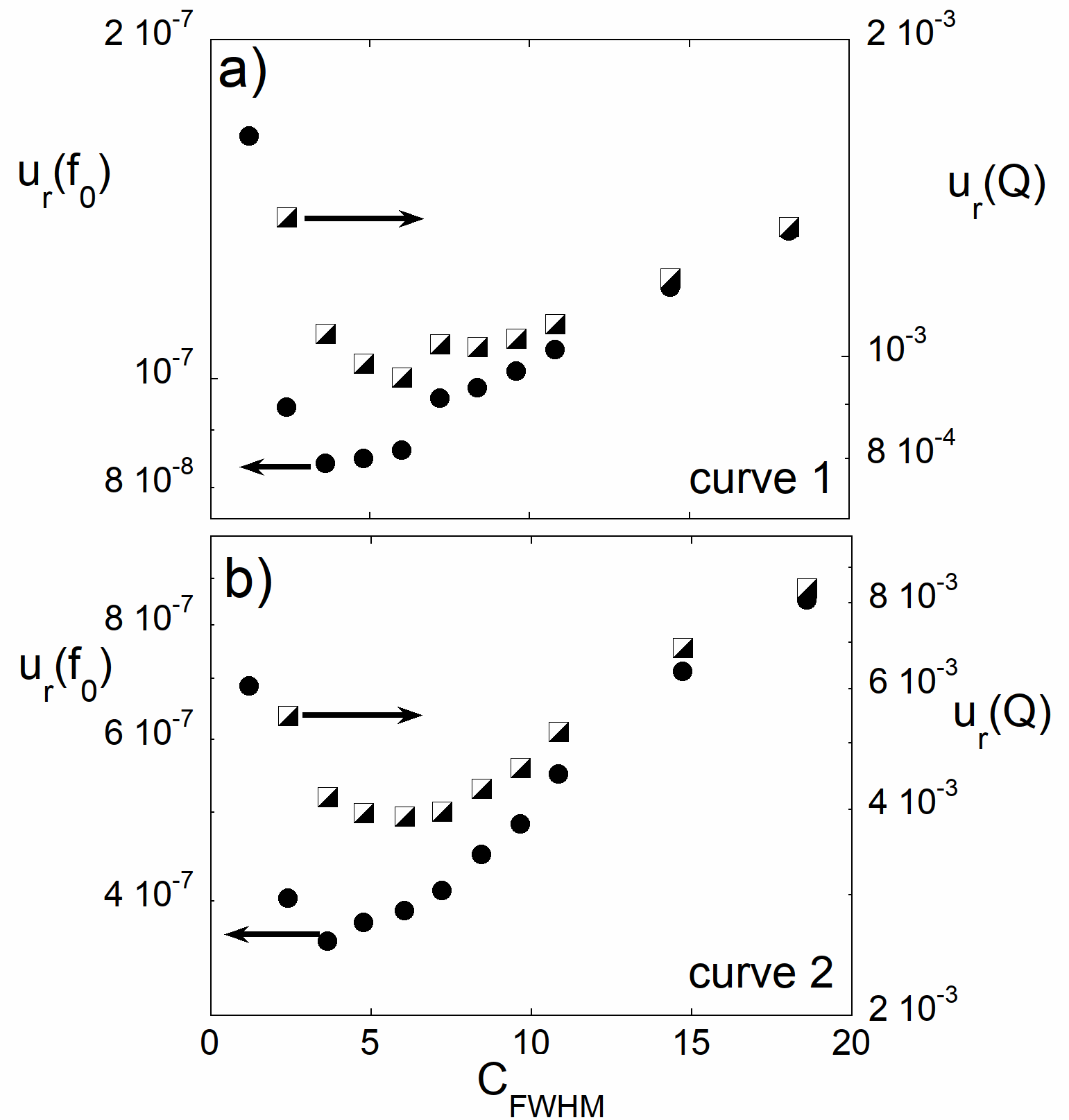}}
\caption{Relative uncertainty on $Q$ and $f_0$ for experimentally measured resonant curves as a function of the span represented as the number of FWHMs.}
\label{fig:uncExp}
\end{figure}

Fig. \ref{fig:uncExp} shows the results of the measurements of the resonant curves as a function of $C_{FWHM}$. 
Even when the resonant curves had evident asymmetry, 
the fit showed that ${S_{C}=(-25+i35)10^{-4}}$ for curve 1 and ${S_{C}=(1.5-i0.85)10^{-4}}$ for curve 2. 

Comparing to the values used in the simulations,this corresponds to a relatively low asymmetry level. 
We obtain that the experimentally determined minimum of the relative uncertainty lays at  $C_{FWHM}=3.7$ for $f_0$, and at $C_{FWHM}=6$ for the $Q$-factor.
This agreement between experimental data and numerical simulations confirms the above recommendation regarding the choice of the optimum measurement span.

\section{Conclusions}
We studied the problem of finding the optimal frequency span for the measurement of a non-ideal resonance curve in order to obtain the minimum uncertainty on the resonator parameters $f_0$ and $Q$.
We addressed the problem both through numerical simulations and through experimental validation. 
The parameter space to be explored was selected within the scenario of limited performance  measurement systems.
By numerically computing a large number of resonant curves with different asymmetries, we determined an optimum span in terms of the number of FWHM, equal to $(4.1 \pm 0.4)$ for resonant frequency and  $(6.1 \pm 0.9)$ for the $Q$-factor. 
These values were confirmed by an experimental study, which has taken advantage of a specifically designed dielectric resonator setup where it was possible to alter in a controlled way the distortion of the resonant curves.

\end{document}